\documentstyle[sprocl,epsfig]{article}

\def\Journal#1#2#3#4{{#1} {\bf #2}, #3 (#4)}

\def\PLB{{\em Phys. Lett.}  B}

\def\PRD{{\em Phys. Rev.} D}

\def\be{\begin{equation}}
\def\ee{\end{equation}}
\def\bea{\begin{eqnarray}}
\def\eea{\end{eqnarray}}

\newcommand{\m}{\rm \,m}

\newcommand{\km}{\rm \,km}

\newcommand{\sr}{\rm \,sr}

\newcommand{\eV}{\rm \,eV}

\newcommand{\lsim}{\lower .5ex\hbox{$\buildrel < \over {\sim}$}} 
\newcommand{\gsim}{\lower .5ex\hbox{$\buildrel > \over {\sim}$}}

\begin{document}
~
\vskip -2.0 cm
{\hskip 5.0 cm \bf University of Bologna Report}

{\hskip 8.0 cm \bf DFUB 01/01}

\title{MACRO AND THE ATMOSPHERIC NEUTRINO PROBLEM}

\author{M. SPURIO for the MACRO collaboration
\footnote{Invited talk at the Third International Workshop on New Worlds in 
Astroparticle Physics 1-3 September 2000, University of the Algarve. Faro, 
Portugal} }

\address{Dipartimento di Fisica dell'Universit\`a and INFN,
        40126 Bologna, Italy\\
        e-mail: spurio@bo.infn.it} 


\maketitle\abstracts{ After a brief presentation of the MACRO detector we 
discuss the updated data on atmospheric muon neutrinos, and 
the interpretation in terms of neutrino oscillations.}

\section{Introduction} 
MACRO is a multipurpose underground detector designed to search for rare events 
in the cosmic radiation. It performs measurements in areas of astrophysics, 
nuclear, particle and cosmic ray physics. In this paper we shall discuss the 
measurement of the atmospheric muon neutrino flux in the energy region from a 
few GeV up to a few TeV, and the interpretation of the MACRO results in terms of 
neutrino oscillations.
The detector has global dimensions of $12\times9.3\times76.5~{\m}^3$ and 
provides a total acceptance to an isotropic flux of particles of $\sim 10,000 
{\m}^2{\sr}$. The total mass is $\simeq 5300 \ $t. A cross section of the 
detector is shown in Fig. \ref{cross_det}a \cite{ncim86}. 
It has three sub-detectors: liquid scintillation counters, limited streamer 
tubes and nuclear track detectors. The mean rock depth of the overburden is 
$\simeq 3700~ $m.w.e.
The average residual energy and the muon flux at the lab depth are $\sim 
310~{GeV}$ and $\sim 1~{\m}^{-2}~{\rm h}^{-1}$, respectively.

\section{MACRO as atmospheric $\nu_\mu$ detector}  
\begin{figure}[tb]
\centerline{
\epsfig{figure=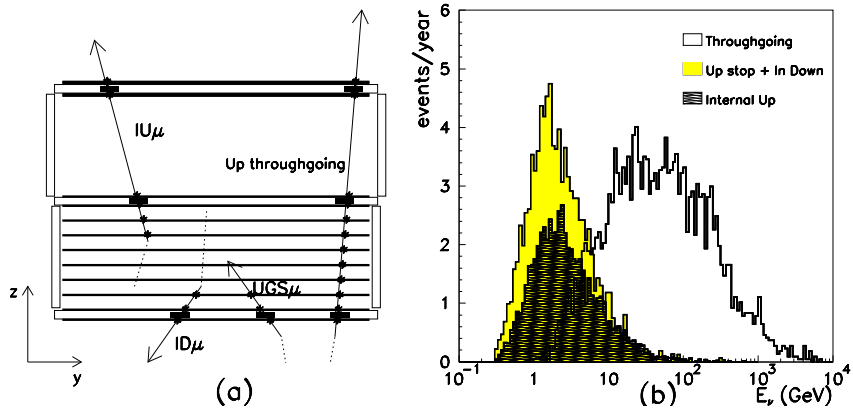,height=4.5cm} }
\vspace{-0.8cm}
\caption {\small a) Event topologies induced by $\nu_\mu$ interactions in and 
around MACRO. The black boxes are scintillator hits.  The track direction and 
versus are measured by the streamer tubes and time of flight. b) Neutrino energy 
distribution for the three event topologies detected by MACRO}
\label{cross_det}
\end{figure}

Primary cosmic rays produce in the upper atmosphere pions and kaons, which 
decay, $\pi \rightarrow \mu \nu $, $K \rightarrow \mu \nu$; and 
$\mu \rightarrow e \nu_e \nu_\mu$. 
Fig. \ref{cross_det}a shows the three different topologies of neutrino events 
detected by MACRO: up throughgoing muons \cite{atmflu}, semicontained upgoing 
muons (IU) and up stopping muons + semicontained downgoing muons (UGS + ID) 
\cite{lownu}. The parent $\nu_\mu$ energy spectra for the three event topologies 
were computed with Monte Carlo (MC) methods, and presented in Fig. 
\ref{cross_det}b. The number of events measured and expected for the three 
topologies is given in Table \ref{tab:macro}.
Sources of background and systematic effects were studied in detail in
\cite{lownu,upgo98} and found to be negligible. The neutrino energies involved 
in all event topologies are such that the $\mu_\nu$ flux is up-down symmetric.

\begin{table}
\begin{center}
\begin{tabular}
{|c|c|c|c|}\hline
              & Events    & Predictions  &R = Data/MC \\
              & selected  & No oscil.    & \\ \hline
Up throughgoing & 723 & 989 & {\bf 0.731}$ \pm 0.028_{st}$\\
              &     &     & $\pm 0.044_{sys}\pm 0.124_{th}$ \\ 
Internal Up   & 135 & 245 & {\bf 0.55} $\pm 0.04_{st} $\\
              &     &     & $\pm 0.06_{sys} \pm 0.14_{th}$ \\ 
UpG Stop +    & 229 & 329 & {\bf 0.70} $\pm 0.04_{st} $   \\
 In Down      &     &     & $\pm 0.07_{sys} \pm 0.18_{th}$ \\ \hline
\end{tabular}
\end {center}
\caption {\small Event summary for the MACRO atmospheric neutrino flux analyses. 
The data correspond to a lifetime of $\sim 5\ y$ with the full detector, after
background corrections [2,3]. The 
ratios R = Data/MC are relative to MC expectations assuming no oscillations.}
\label{tab:macro}
\end{table}

\begin{figure}
 \vspace{-2.5cm}
 \begin{center}
  \mbox{ \epsfysize=5.5cm
         \hspace{-1.2cm}
         \epsffile{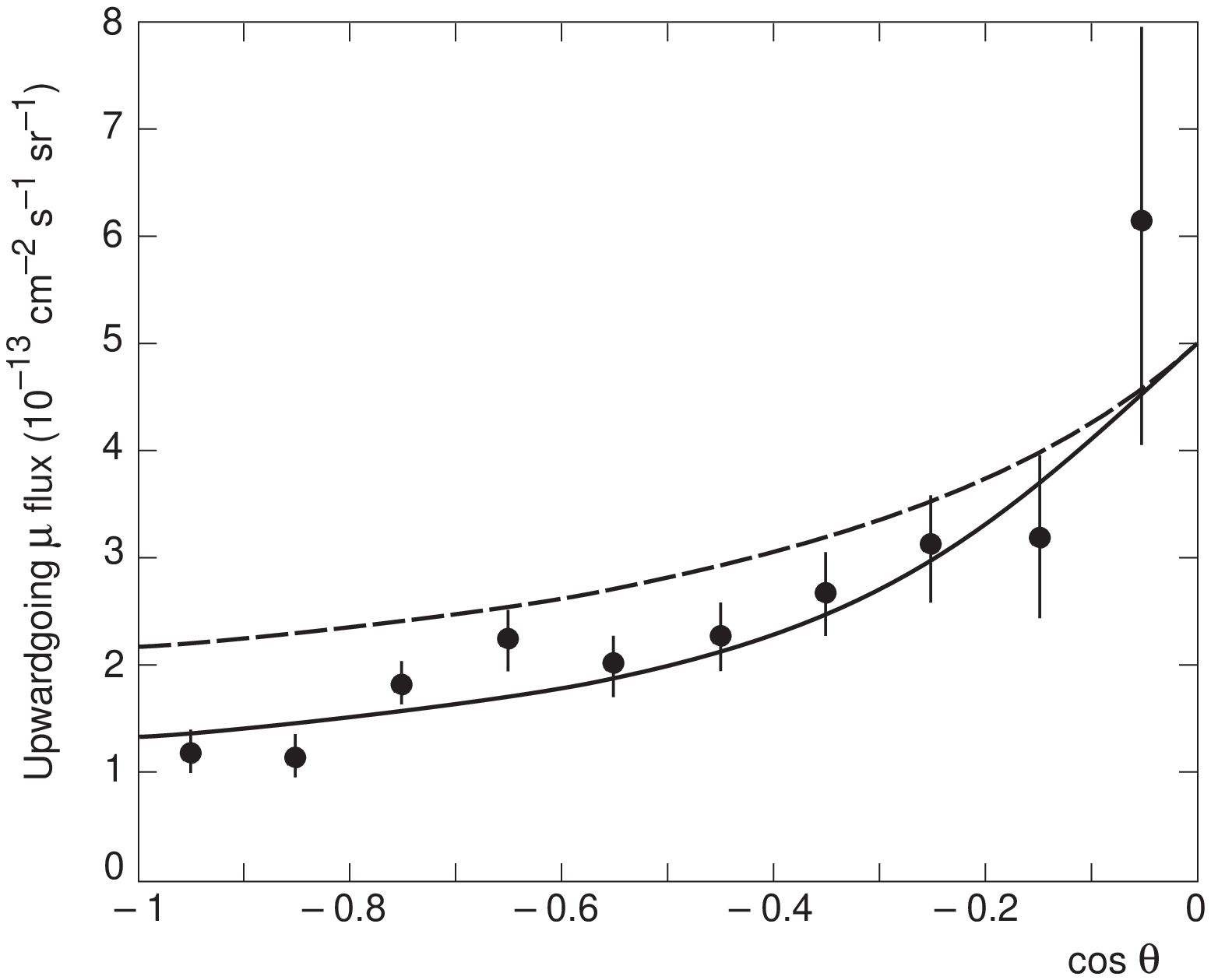}\hspace{0.1cm}
         \epsfysize=5.5cm
         \vspace{3.5cm}
         \epsffile{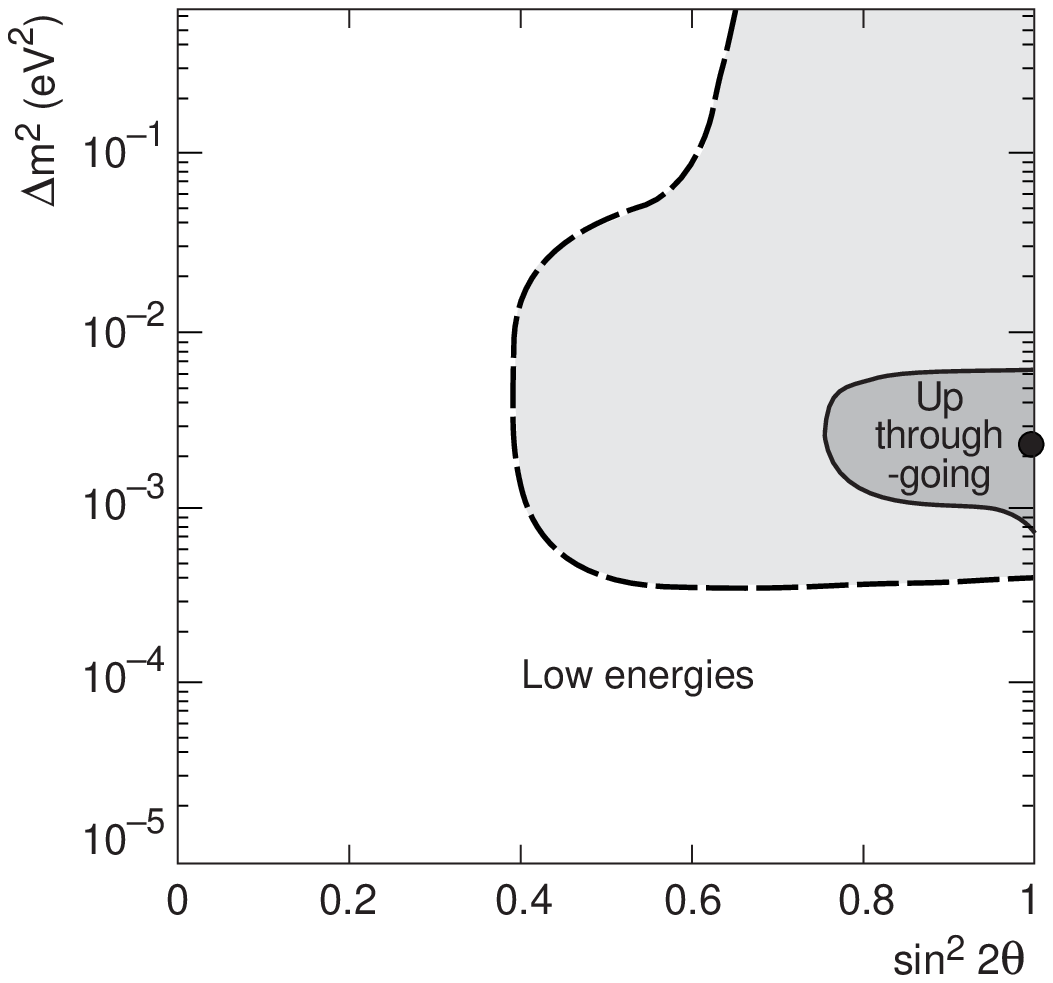} }
 \end{center}
\caption{\label{flux}\small  (a) Measured flux (points) of the up throughgoing 
muons plotted vs. zenith angle $\Theta$. The solid line is the fit to an 
oscillated muon flux, obtaining maximum mixing and $\Delta m^{2} = 0.0025$ 
eV$^{2}$. (b) Confidence regions for $\nu_\mu \rightarrow \nu_\tau$ oscillations 
at the 90 \% CL, obtained from MACRO high energy [2] and low energy [3] data.} 
\end{figure}

{\bf High energy sample.}
The {\it up throughgoing muons} come from $\nu_\mu$'s interacting in the rock 
below the detector; they are identified using the streamer tube system (for 
tracking) and the scintillator system (for time-of-flight (ToF) measurement).
A rejection factor of at least $10^5$ is needed in order to separate the up-
going muons from the large background of  down-going muons 
\cite{atmflu}. The $\nu_{\mu}$ median energy is $\overline E_\nu \sim \ 50\ 
GeV$, and muons with $E_\mu > 1\ GeV$ cross the whole detector. 
Fig. \ref{flux}a shows the zenith angle distribution of the measured flux of up 
throughgoing muons.
The data have been compared with MC simulations \cite{atmflu}; the expected 
angular distributions for no oscillations is shown in Fig. \ref{flux}a as a 
dashed line, and for a $\nu_\mu \rightarrow \nu_\tau$ oscillated flux with
$\sin^2 2\theta =1$ and  $\Delta m^2= 0.0025\ \eV ^2$, as a solid line.
The total theoretical uncertainty on the expected muon flux, adding in 
quadrature the errors from neutrino flux, cross section and muon propagation, is 
17 \%; it is mainly a scale error that does not change the shape of the angular 
distribution. The ratio of the observed number of events to the expectation 
without oscillations is given in Table \ref{tab:macro}.

\begin{figure}[tbh]
\centerline{
\epsfig{figure=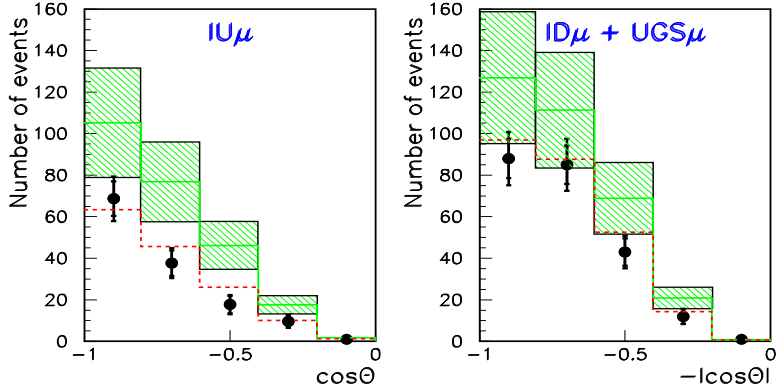,height=4cm} }
\vspace{-0.8cm}
\caption {\small  
Measured (points) and expected number of low energy muon neutrino events versus 
zenith angle. IU: up semicontained. ID+UGS: up stopping plus down semicontained 
muons. The shadowed regions are the predictions without oscillations; the dotted 
lines are the predictions assuming neutrino oscillations with the parameters 
obtained from the up throughgoing sample.}
\label{dist_ang}
\end{figure}

{\bf Low energy samples.}
The {\it upgoing semicontained muons} (IU, in Fig. \ref{cross_det}a) come from 
$\nu_\mu$ interactions inside the lower apparatus. Since two scintillation 
counters are intercepted, the ToF is applied to identify the upward going muons. 
The average parent neutrino energy for these events is 4.2 GeV. The {\it upgoing 
stopping muons} (UGS) are due to external $\nu_\mu$ interactions yielding 
upgoing muon tracks stopping in the detector; the {\it semicontained downgoing 
muons} (ID) are due to $\nu_\mu$ induced downgoing tracks with vertices in the 
lower MACRO. The events were selected by means of topological criteria; the lack 
of time information prevents to distinguish the two sub samples, for which an 
almost equal number of events is expected. The average neutrino energy for these 
events is $\simeq 3.5\ GeV$.
The number of events and the angular distributions are compared with the 
predictions \cite{lownu} in Fig. \ref{dist_ang} and Table \ref{tab:macro}.

\section {Interpretation in terms of neutrino oscillations}

We interpreted the reduction on the detected number of events (Table 1) and the 
deformation of the zenith angle distribution of the up-throughgoing events (Fig. 
2a) as a consequence of $\nu_\mu$ disappearance. 

The weak flavor eigenstates ($\nu_e$, $\nu_\mu$, $\nu_\tau$) are relevant for 
the production of atmospheric neutrinos, via $\pi \rightarrow \mu \nu_\mu$, 
K$\rightarrow \mu \nu_\mu$ and  $\mu \rightarrow e \nu_e \nu_\mu$ decays.
For massive neutrinos one has to consider the mass eigenstates ($\nu_1$, 
$\nu_2$, $\nu_3$) which are relevant for propagation. The flavor eigenstates are 
linear combinations of the mass eigenstates, \( \nu_l = \sum_{m=1}^{3} U_{lm} 
\nu_m\).
In the simplest two flavors oscillations scenario, only one mixing angle $\theta 
$ is involved, and the rate of $\nu_\mu$ disappearance is given by 
\begin{equation} P(\nu_\mu \rightarrow \nu_\mu) \simeq 1 - sin^2 2\theta \, 
sin^2 (1.27 \Delta m^2  L / E_\nu) \end{equation}
$\Delta m^2 = {m^2}_{\nu_3} - {m^2}_{\nu_2} $, {\it L} = distance (in km) from 
the  decay point to the interaction point, $E_\nu$ is the neutrino energy, 
in GeV.

This simple relation could be modified when neutrinos propagate
with a flavor-dependent interaction through matter \cite{ronga}.
This case is important, because the {\it matter effect} could help to 
discriminate between $\nu_\mu$ oscillation in different neutrino channels. The 
matter effect in the Earth could enhance (in a range of values of the 
oscillations parameters) the $\nu_\mu$ disappearance effect 
through "resonances" 
(MSW effect). This happens only for $\nu_\mu \rightarrow \nu_e$ and 
for $\nu_\mu \rightarrow \nu_s$ channels ($\nu_s = sterile\ neutrino$), while 
for $\nu_\mu \rightarrow \nu_\tau$ there 
is no matter effect ($\nu_\mu$ and $\nu_\tau$ have 
the same weak potential). 

In the scenario described by eq. 1, relatively fewer up throughgoing muons are 
expected near the vertical $(cos\Theta=-1)$ than near the horizontal 
$(cos\Theta=0)$, due to the longer path length of neutrinos from production to 
observation, in a wide range of $\Delta m^2$ values.
The probabilities to obtain the number of events in Table 1 and the angular 
distribution observed in Fig. 2a have been calculated $vs.$ oscillation 
parameter values. 
For $\nu_\mu \rightarrow \nu_\tau$ oscillations the maximum probability is 57 
\%, for $\Delta m^2 = 0.0025\ \eV ^2$,  $\sin^2 2\theta = 1$ (combination from 
the shape of the angular distribution and the reduction in the number of 
events). Fig. \ref{flux}b  shows the 90 \% CL regions for $\nu_\mu \rightarrow 
\nu_\tau$. 

The probability for no-oscillations is 0.4\%. The maximum probability for 
$\nu_\mu \rightarrow \nu_{s}$ oscillations is 15\%. Another way to 
discriminate between the $\nu_\mu$ oscillation in $\nu_\tau$ or in  $\nu_{s}$ is 
to study the angular distribution in two bins. The statistical significance is 
higher \cite{lipariste} with respect to the 10-bins fit of the angular 
distribution. 
From this ratio, the $\nu_\mu \rightarrow \nu_s$ hypothesis is disfavored at 
level of 2.5$\sigma$ compared to $\nu_\mu \rightarrow \nu_\tau$\cite{ronga}.


From our results on the high energy sample, a $\sim$ 50\% reduction in the flux 
of the up stopping events (UGS) as well of semicontained upgoing muons (IU) is 
expected, with no distortion in the shape of the angular distribution. No 
reduction is instead expected for the semicontained downgoing events (ID, coming 
from neutrinos with path lengths of $\sim 15 \km$). 
This is what is observed in Fig. \ref{dist_ang}.
The allowed region in the $\Delta  m^2$,  $\sin^2 2\theta $ plane is shown in
Fig. \ref{flux}b.\par

\section{Up throughgoing muon energy separation}
The up throughgoing muon sample is induced by atmospheric $\nu_\mu$ in a wide 
energy range (for 90\% of events $5 \lsim \ E_\nu \ \lsim 700$ 
GeV, and $2 \lsim \ E_\mu \ \lsim 300$ GeV, see Fig. 1b).
If the observed event reduction is due to neutrino oscillations,
the first minimum of eq. 1 occurs at $E_\nu \simeq 10-30 \ 
GeV$ (for neutrinos from the nadir direction and oscillation parameters 
in the range of Fig. \ref{flux}b). 
Thus, the $\nu_\mu$ disappearance 
mechanism works on (relatively) lower energy neutrinos.
As a consequence, we expect (from MC) a $\sim$ 
50\% reduction of the up throughgoing muons with $E_\mu \lsim 10 \ GeV$, and 
only a $\sim$ 20\% reduction for muons with $E_\mu \gsim 10 \ GeV$.

Muons traversing MACRO are deflected by many small-angle scatters, the bulk of 
which are Coulomb scatterings from the nuclei and the atomic electrons 
(multiple 
Coulomb scattering). As an effect, relatively low energy up throughgoing muons 
are (slightly) deviated from a straight-line direction inside the detector. 
We evaluated that this 
deviation is of the order of few centimeters, up to $E_\mu \sim 10 \ GeV$, 
and it exceed the intrinsic spatial resolution of the MACRO streamer tube cells.

A selection cut, based on the deviation of the muon sagitta from the straight 
line when the muon cross the whole detector, was defined. 
Each up throughgoing muon is classified as: 
{\it Low}= muon with a deviation from the straight
line (from MC, lower energy muons with median $E^{Low}_\mu= 11\ GeV$); 
{\it High}= muon with no measurable deviation 
(muons with median energy $E^{High}_\mu = 52\ GeV$). The remaining events 
with ambiguous deviation, and
which does not satisfy the high or the low energy cuts, 
were classified as {\it Medium}. From MC, 
they have an intermediate energy $E^{Medium}_\mu=33\ GeV$. 
The same cuts were then
applied to the simulated neutrino-induced events, and the results compared.
To avoid systematic differences from data and MC,
a detailed check of the detector simulation was  
performed. We used
a large sample of measured and simulated 
downward going atmospheric muons, and we required  that
the distributions, used to define the selection cut, match together. 
The (preliminary) results 
are presented in Fig. 4. It is included also the point corresponding to the 
partially contained IU events.   Only 87\% of IU events have been induced by
$\nu_\mu$ CC interactions, the remaining by $\nu_e$ and NC interactions.

Fig. 4 shows that lower energy events are more suppressed than higher 
energy ones. In a similar way, Fig. 2a shows that neutrinos with 
a longer path length are more
suppressed; both effect are well explained by neutrino oscillation (eq. 1),
with the oscillation parameter presented in Fig. 2b.

\begin{figure}[tb]
\centerline{
\epsfig{figure=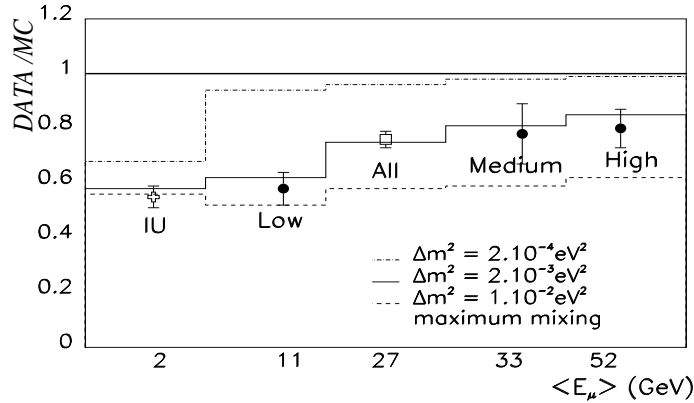,height=5cm} }
\vspace{-0.8cm}
\caption {\small Ratio (detected events/expected events) vs. median muon energy 
for the whole sample of 723 up throughgoing muons (All), 
and for the sub sample with large (Low, 79 events) and
small (High, 122 events) deviation from a straight line. 
It is also included the ratio for the events with intermediate 
energy (Medium, 49 events). The point IU corresponds to the
135 partially contained upgoing events. For all samples, 
the $\nu_\mu$ path length is $\sim 10^4\ km$. 
The points are plotted with equal distance in abscissa, to guide the eye.
The full line (data/MC=1) is 
the expectation without neutrino oscillations, for which
a 17\% total uncertainty is associated (25\% for the IU point). 
The dashed lines are the expected reductions in case of 
$\nu_\mu \rightarrow \nu_\tau$ oscillations with maximum mixing 
and three values of $\Delta m^2$.}
\label{musca}
\end{figure}

I acknowledge all the colleagues of the MACRO collaboration, in particular
D. Bakari and Y. Becherini.



\section*{References}
\begin{thebibliography}{99}

\bibitem[1]{ncim86} S. Ahlen et al., Nucl. Instr. Meth. A324(1993)337.  
\bibitem[2]{atmflu}M. Ambrosio et al., Phys. Lett. B434(1998)451; 

Phys. Lett.  B357(1995)481.
\bibitem[3]{lownu}M. Ambrosio et al.,  Phys. Lett. B478(2000)5;

M. Spurio, Nucl.Phys.Proc.Suppl.85:37-43,2000 (hep-ex/9908066).
\bibitem[4]{upgo98} M. Ambrosio et al., Astropart. Phys. 9(1998)105.
\bibitem[5]{ronga} F. Ronga, Nucl.Phys.Proc.Suppl.87:135,2000 
(hep-ex/0001058), and references therein.
\bibitem[6]{lipariste} P. Lipari and M.Lusignoli, Phys. Rev. D57, 3842(1998).

\end{thebibliography}
\end{document}